\documentclass[pdflatex,sn-mathphys]{sn-jnl}

\usepackage[T1]{fontenc}

\jyear{2022}%

\theoremstyle{thmstyleone}%
\theoremstyle{thmstyletwo}%

\theoremstyle{thmstylethree}%

\begin{document}

\title[Night Sky Brightness]{Night Sky Brightness Measurement, Quality Assessment and Monitoring}

\author*[1,2]{\fnm{John~C.} \sur{Barentine}}\email{john@darkskyconsulting.com}
\affil*[1]{\orgname{Dark Sky Consulting, LLC}, \orgaddress{\street{PMB 237, 9420 E. Golf Links Rd., Ste 108}, \city{Tucson}, \postcode{85730-1317}, \state{AZ}, \country{USA}}}
\affil[2]{\orgdiv{Consortium for Dark Sky Studies}, \orgname{University of Utah}, \orgaddress{\street{375 S 1530 E, RM 235 ARCH}, \city{Salt Lake City}, \postcode{84112-0730}, \state{UT}, \country{USA}}}

\abstract{Ground-based optical astronomy necessarily involves sensing the light of astronomical objects along with the contributions of many natural sources ranging from the Earth's atmosphere to cosmological light. In addition, astronomers have long contended with artificial light pollution that further adds to the `background' against which astronomical objects are seen. Understanding the brightness of the night sky is therefore fundamental to astronomy. The last comprehensive review of this subject was nearly a half-century ago, and we have learned much about both the natural and artificial night sky since. This Review considers which influences determine the total optical brightness of the night sky; the means by which that brightness is measured; and how night sky quality is assessed and monitored in the long term.}


\maketitle

\section{Introduction}\label{sec:intro}

Environmental pollution caused by artificial light at night, commonly known as ``light pollution,'' is a source of significant known and suspected hazards~\cite{Gaston2015,Falchi2018,Green2022}. Light pollution now touches every continent except Antarctica~\cite{Falchi2016} and yields steadily increasing environmental pressure.~\cite{Gaston2014,Gaston2021} Of the world population, more than 80\% of all people and more than 99\% of people in the U.S. and Europe live in places where the brightness of the night sky is elevated due to light pollution.~\cite{Falchi2016} Both the extent to which the indication of artificial light appears in satellite remote sensing data and the quantity of emitted light have increased globally on average by about two percent per year in recent years.~\cite{Kyba2017} The spatial variance of anthropogenic light is large,~\cite{Falchi2019} and both lit area and quantity of light are stable or decreasing in only a handful of countries.~\cite{Kyba2017}

Light pollution manifests itself both as a presence on the ground and in the night sky. On the ground, we perceive its effects directly in forms such as glare and light trespass,~\cite{Sim2016} and indirectly in threats to human and wildlife biology,~\cite{Schroer2016,Svechkina2020,Boyce2021} public safety,~\cite{Wanvik2009,Marchant2020} and energy security.~\cite{Kyba2014,SchulteRomer2019} In the night sky, light pollution yields skyglow, a condition in which artificial light directed upward is scattered back to the ground where it obscures our view of the stars.~\cite{Cinzano2020} The brightness of the night sky, relative to its assumed `pristine' state absent anthropogenic light pollution, is related to the amount of artificial light emitted on the ground,~\cite{Garstang1986,Kocifaj2014} so night sky quality is often taken as a proxy for the conditions affecting natural darkness on Earth.~\cite{Duriscoe2013,Hung2021}

Managing the resource of natural nighttime darkness involves understanding its nature and the influences that threaten its integrity, which in turn require knowing the initial state of the resource and how that state evolves with time.~\cite{Barentine2019} Ground-based optical astronomy depends crucially on this knowledge in order to extract the maximum information value from the cosmic light that our telescopes and instruments collect.~\cite{Green2022} Understanding the sources of light controlling the brightness of the night sky and how those components change on various timescales is therefore a fundamental concern to astronomy. 

The factors that determine night sky brightness were last comprehensively reviewed in the literature almost 50 years ago.~\cite{RoachGordon1973} In the time since, research combining elements of astronomy, atmospheric science and space physics has revealed a more complete picture of those factors. The deeper understanding of this phenomenon that results has implications for both how modern ground-based astronomical observatories are designed and operated as well as the outdoor lighting policies and practices that best support observatory site protection.

This Review considers which factors determine the brightness of the night sky (§~\ref{sec:nsb-components}); how sky brightness is quantified (§~\ref{sec:sensing-monitoring}); relevant units of measurement (§~\ref{sec:units}); and how night sky quality is classed, compared and ranked (§~\ref{sec:ranking}). Throughout the Review we refer to two acronyms frequently, mirroring their use in the literature: ``night sky brightness'' (NSB), and ``artificial light at night'' (ALAN).

%
%
\section{The natural and artificial night sky}\label{sec:nsb-components}

\subsection{As night falls: setting the stage}\label{subsec:as-night-falls}
\begin{figure}[tbph]
\begin{center}
\includegraphics[width=88mm,height=111mm,angle=0]{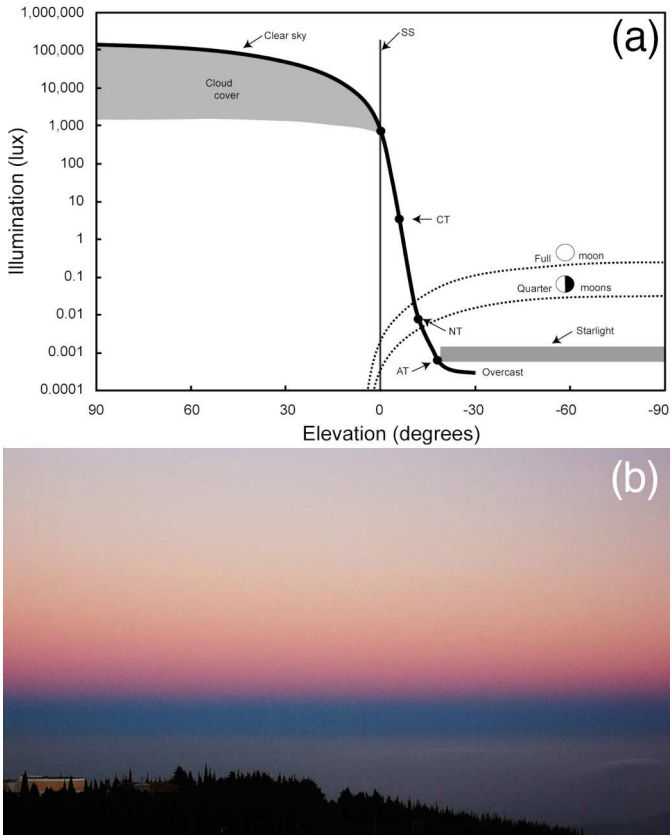}
\caption{Natural light in the sky from day to night. (a) Natural outdoor illumination levels adapted from~\cite{Beier2006}. Horizontal illuminance is shown on the ordinate, while the abscissa shows the altitudes of the Sun (heavy solid line) and Moon (dotted lines). SS = sunset, CT = civil twilight, NT = nautical twilight, AT = astronomical twilight. Figure courtesy of T. Longcore. (b) The anti-twilight arch appears just above the Earth’s shadow at dusk. Photo by G. Donatiello.\label{twilight-illumination}}
\end{center}
\end{figure}

From sunset to the onset of astronomical darkness, the brightness of the sky at the zenith decreases by a factor of about $4\times10^{5}$ (Figure~\ref{twilight-illumination}a). The time required to complete this transition varies with season and latitude; in the tropics it can take place in as little as 72 minutes. After sunset, the Sun continues to directly illuminate the atmosphere for some time. This period is called twilight, during which sky brightness is dominated by the scattering of sunlight illuminating the atmosphere at progressively higher altitudes.

Since the amount of scattered sunlight depends on the number of scatterers along the direction of travel of light rays, and the density of scatterers decreases with height above the surface of the Earth, the brightness of the twilight sky drops rapidly as the Sun descends further below the horizon. First-order scattering of light dominates sky brightness from sunset to until about the end of civil twilight, when the solar altitude reaches $-6^{\circ}$. As the angle increases, second- and higher-order scattering become important.~\cite{Belikov1996}

The color of the twilight sky is generally the same blue as the daytime sky except for contributions resulting from a combination of Rayleigh scattering and absorption by ozone (O$_{3}$) molecules whose concentration in the stratosphere peaks at altitudes of $30-35$ kilometers.~\cite{Adams1974} When there are unusual quantities of particulates in the atmosphere, such as following significant wildfire events and volcanic eruptions, scattered sunlight is subject to additional `de-blueing', resulting in vivid colors. The same circumstances are known to cause significant variations in atmospheric extinction, affecting astronomical photometry.~\cite{Burki1995}

Opposite the setting Sun in the sky, the anti-twilight arch rises (Figure~\ref{twilight-illumination}b). Its warm tones represent the enhancement of backscattered sunset light de-blued by its passage through the atmosphere at near-grazing angles. It reaches a maximum width of between $3-6^{\circ}$ above the anti-solar point; below the arch, a deep blue tinged with purple indicates the projection of the Earth's shadow onto the twilight sky.

Direct illumination of the atmosphere reaches the local zenith when the altitude of the Sun is approximately $-6^{\circ}$ and second-order scattering begins to become important. Around this time the lower boundary of direct solar irradiation reaches 35 km altitude where it excites a layer of neutral sodium atoms to the $^{2}P^{0}$ state, causing them to emit in the 589.0/589.6-nm ``D'' lines of Na \textsc{i}. By the time solar angle reaches $-6.5^{\circ}$, irradiation is reduced to the point where the intensity of the emission equals that of the corresponding absorption lines in the solar spectrum. The episode is so relatively short in duration that it has been described as the ``sodium flash.''~\cite{Hunten1964}

\begin{figure}[tbph]
\begin{center}
\includegraphics[width=88mm,height=176mm,angle=0]{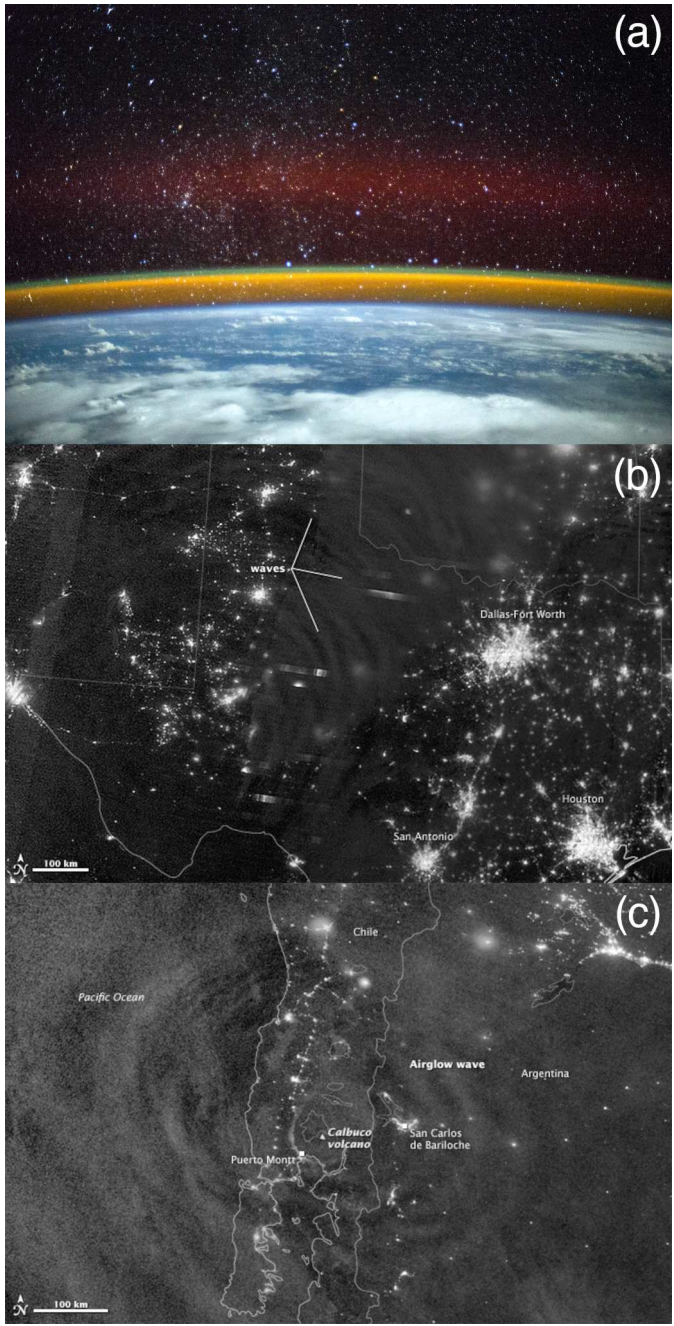}
\caption{Night airglow phenomena seen from Earth orbit. (a) Emitting layers of neutral sodium (yellow; 80-100 km) and oxygen (green, 100 km; red, 200-300 km) atoms seen in an oblique view of the Earth’s limb from low- earth orbit near the terminator. NASA photo ISS042-E-037847 taken by European Space Agency astronaut Samantha Cristoforetti. (b) Structured airglow observed above thunderstorms in west Texas, U.S., originating along a dry line on 15 April 2012. (c) Airglow waves emanating from the site of the eruption of the Calbuco volcano in southern Chile on 22 April 2015. Images by Jesse Allen (NASA Earth Observatory).\label{airglow}}
\end{center}
\end{figure}
As twilight progresses, a similar ``oxygen flash'' happens as neutral oxygen atoms at altitudes from 
$200-300$ km are directly irradiated by the Sun, yielding emission in the $^{1}S{\rightarrow}^{3}P$ 630/636-nm lines of O \textsc{i}. This is thought to contribute to the purplish hues of the sky during late twilight along with molecular scattering of sunlight at high altitudes and de-blued illumination of the stratosphere and the troposphere.~\cite{Lee2003} Since emission in the O \textsc{i} doublet is radiatively excited, the lines largely fade out after the onset of astronomical darkness. 

However, a chemical process excites the same transitions on the night side of Earth, particularly at low magnetic latitudes, resulting in red oxygen ``airglow'' that varies in intensity and sky distribution through the night in the tropics.~\cite{Tinsley1973,Slanger2006} Oxygen atoms at lower altitudes are collisionally excited and emit characteristically in the forbidden $^{1}S{\rightarrow}^{1}D$ transition at 577 nm. Resulting O \textsc{i} emission persists throughout the night.~\cite{Bates1978} Figure~\ref{airglow}a shows the stratification of sodium and oxygen airglow toward the limb of the Earth as seen from low-earth orbit. This phenomenon is discussed further in Section~\ref{subsubsec:airglow-aurorae}.

Direct illumination of the entire atmosphere above an observer continues until the Sun reaches an altitude of $-12^{\circ}$ (nautical dusk). From here to a solar altitude of $-18^{\circ}$, only the uppermost parts of the atmosphere seen in the direction of the horizon are still directly illuminated by the Sun. Once the Sun descends below $-18^{\circ}$, direct illumination ceases and astronomical darkness begins. The sequence of events described here unfolds in reverse on the morning side of night, from the point when the Sun again reaches an altitude of $-18^{\circ}$ until sunrise.

%
%
\subsection{Sources of natural light in the night sky}\label{subsec:natural-light}
The total brightness at any point on the night sky is the sum of contributions from both natural and artificial sources, each of which is a function of azimuth ($\alpha$) and altitude ($\gamma$). Some sources are also a function of time ($t$),~\cite{Patat2008} so in general:
\begin{equation}
B_{sky}(\alpha, \gamma, t) = \sum_{i=1}^{N} B_i(\alpha, \gamma, t)
\end{equation}

The natural, optical-light components of NSB have been reviewed in several publications,~\cite{RoachGordon1973,Leinert1998,Duriscoe2013} and are summarized in Tables~\ref{sources-table-atmospheric} and~\ref{sources-table-astronomical}. These components are divided into two sources: those that originate in or near the Earth's atmosphere and those that originate in the space beyond the Earth's atmosphere. Because the boundary between the atmosphere and outer space is not well defined, we include the Earth's magnetic environment as part of `atmosphere' such that, e.g., aurorae are not considered an astronomical phenomenon. Kocifaj et al.~\cite{Kocifaj2021b} recently showed that artificial objects orbiting the Earth are a non-trivial contributor of diffuse light to the night sky as seen from the ground. We have included a value for this effect in Table~\ref{sources-table-atmospheric}. On the other hand, distinctly terrestrial sources of natural light at night, such as wildfire, lightning and bioluminescence, are typically neglected in NSB calculations because they are regarded as insignificant contributors except on highly local spatial scales. 

%
%
\renewcommand*{\thefootnote}{\alph{footnote}}
\begin{sidewaystable}
\sidewaystablefn%
\begin{center}
\begin{minipage}{\textheight}
\caption{Sources of natural optical light in the night sky originating in and near the Earth's atmosphere. For each component (column 1), the angular extent of the light in the sky (column 2) is given along with its average surface brightness (column 3), the factors that influence its extent and/or brightness (column 4), and its physical cause (column 5).}\label{sources-table-atmospheric}
\begin{tabular*}{\textheight}{@{\extracolsep{\fill}}lccll@{\extracolsep{\fill}}}
\toprule%
\textbf{Component}  & \textbf{Sky} & \textbf{Average brightness} & \textbf{Dependencies} & \textbf{Physical origin} \\
 & \textbf{distribution} & (S$_{10}$(vis))\footnotemark[1] &  &  \\
\midrule
Airglow & Variable & 50 (continuum)\footnotemark[2] & Local time, elevation, & 
Ambient excitation of atoms  \\
 & (extended) & 145-270 (line)\footnotemark[3] & latitude, season, solar activity, & and molecules in the upper \\
  & & & wavelength & atmosphere \\
\hline
Aurora & Variable & IBC Class\footnotemark[4] I: 70\footnotemark[5] & Local time, magnetic latitude, & Excitation of atoms and \\
 & (extended) & IBC Class II: 700 & solar cycle phase, solar cycle & molecules in the upper \\
  & & IBC Class III: 7000 & intensity, wavelength & atmosphere by charged particles \\
  & & IBC Class IV: 70000 & & \\
\hline
Atmospheric & $\phi\gtrsim80^{\circ}$, & $\leq$50\footnotemark[7]  & Atmospheric aerosol & Multiple scattering of \\
diffuse light\footnotemark[6] & $0^{\circ}\leq\theta\leq360^{\circ}$ & & optical depth & all sources of light \\
 & (horizon coordinates) & & & in the sky \\
\hline
Space objects\footnotemark[8] & Extended\footnotemark[9] & $>$15\footnotemark[10] & Number, albedo and size & Reflected sunlight \\
 & & & distributions of & \\
 & & & space objects & \\
\botrule
\end{tabular*}
\footnotetext[a]{S$_{10}$(vis) is a linear unit equal to the surface brightness of a star whose visual magnitude is +10 and whose light is distributed over one square degree. In SI units, 1 S$_{10}$(vis) $\approx$ $1.04\times10^{-6}$ cd m$^{-2}$.}
\footnotetext[b]{\cite{Sternberg1972}}
\footnotetext[c]{\cite{RoachGordon1973}}
\footnotetext[d]{The large range in auroral brightness is rated from zero to four on a base-ten logarithmic scale (the International Brightness Coefficient, or IBC; ~\citealp{Seaton1954,Hunten1955}).}
\footnotetext[e]{All auroral brightnesses are drawn from~\cite{Chamberlain1961}.}
\footnotetext[f]{\cite{Hong1998,Hong2002,Kwon2004}.}
\footnotetext[g]{The ADL reaches a maximum at very large zenith angles ($\sim90^{\circ}$).~\cite{Kwon1989}}
\footnotetext[h]{Defined in~\cite{Kocifaj2021b} to include both satellites and space debris.}
\footnotetext[i]{In addition to an overall diffuse glow, NSB contributions from space objects may be higher along the paths of certain orbital shells across the sky. See, e.g.,~\cite{Bassa2022}.}
\footnotetext[j]{Estimated value as of mid-2019.}
\end{minipage}
\end{center}
\end{sidewaystable}

%
%
\renewcommand*{\thefootnote}{\alph{footnote}}
\begin{sidewaystable}
\sidewaystablefn%
\begin{center}
\begin{minipage}{\textheight}
\caption{Sources of natural optical light in the night sky originating beyond the Earth's atmosphere. The order and contents of the columns are the same as in Table~\ref{sources-table-atmospheric}.}\label{sources-table-astronomical}
\begin{tabular*}{\textheight}{@{\extracolsep{\fill}}lccll@{\extracolsep{\fill}}}
\toprule%
\textbf{Component}  & \textbf{Sky} & \textbf{Average brightness} & \textbf{Dependencies} & \textbf{Physical origin} \\
 & \textbf{distribution} & (S$_{10}$(vis)) &  &  \\
\midrule
Zodiacal & $\lambda-\lambda_{\odot}$, $\beta$ & $\beta=\pm90^{\circ}$: 78\footnotemark[2] & Season, latitude & Sunlight scattered by \\
light & (ecliptic coordinates)& ($140^{\circ}$,$0^{\circ}$): 164 & & interplanetary dust \\
  & & ($90^{\circ}$,$0^{\circ}$): 250 & & ($0^{\circ}<\phi\leq180^{\circ}$) \\
 & & ($60^{\circ}$,$0^{\circ}$): 500 & & \\
 & & ($30^{\circ}$,$0^{\circ}$): 2330 & & \\
\hline
Gegenschein & $175^{\circ}\lesssim\lambda-\lambda_{\odot}\lesssim185^{\circ}$; & $40-205$\footnotemark[3] & & Sunlight backscattered by \\
 & $-5^{\circ}\lesssim\beta\lesssim+5^{\circ}$ & & & interplanetary dust \\
 & & & & ($\phi=0^{\circ}$)\\
\hline
Integrated & $\ell,b$ (galactic & $b=\pm90^{\circ}$: $25-30$\footnotemark[4] & Season, latitude & Unresolved stars \\
starlight & coordinates) & $b=0^{\circ}$: $100-260$\footnotemark[5] & & in the Milky Way \\
\hline
Diffuse & $\ell,b$ & $\leq66$\footnotemark[6] & &Starlight scattered by \\
galactic light & & & & interstellar dust \\
\hline
Extragalactic & $\ell,b$ & $\leq$2\footnotemark[7] & Cosmological model, & Unresolved \\
background & & & cosmological redshift & galaxies\footnotemark[8] \\
light & & & & \\
\botrule
\end{tabular*}
\footnotetext[a]{All zodiacal light brightnesses are as reported in~\cite{Leinert1998}.}
\footnotetext[b]{\cite{Tanabe1965,RoachGordon1973,Leinert1975,James1997,Buffington2009}.}
\footnotetext[c]{\cite{Nawar2010}.}
\footnotetext[d]{\cite{Elsasser1960}.}
\footnotetext[e]{At 4250 \AA.~\cite{WittLillie1973,Toller1981}.}
\footnotetext[f]{\cite{Dube1977}.}
\footnotetext[g]{`Diffuse Cosmic Optical Background'; \cite{Lauer2021}.}
\footnotetext[h]{Lauer et al.~\cite{Lauer2022} report ``a flux component of unknown origin'' of 8.06$\pm$1.92 nW m$^{-2}$ sr$^{-1}$ in New Horizons Long-range Reconnaissance Imager (LORRI) measurements at $\lambda=0.608$ $\mu$m after subtracting the estimated contribution from the integrated light of external galaxies.}
\end{minipage}
\end{center}
\end{sidewaystable}

Natural sources of light in the night sky range over several orders of magnitude in both surface brightness and wavelength.~\cite{Leinert1998} The total natural NSB, generally quoted for the zenith, is the sum of the astronomical sources with an allowance for quiescent, (pseudo-)continuum airglow, but not including other transient terrestrial sources like aurora. This averages about 205 S$_{10}$(vis) or $\sim22.0$ m$_{V}$ mag arcsec$^{-2}$ ($\sim2\times10^{-4}$ cd m$^{-2}$).\footnote{Assuming the ecliptic pole were viewed at the zenith and with no contribution from aurora or airglow. See the further discussion of the minimum brightness of the night sky in Section~\ref{subsec:subjective}.} This value is itself proposed as a unit of measurement, called one Night Sky Unit (NSU) or one ``sky''.~\cite{Kyba2015}

Note that the brightnesses of all natural light phenomena are quoted for ``clear air'' conditions at typical atmospheric optical depth ($\tau$) values near mean sea level. NSB values $<1$ NSU reported in the literature result from light losses due to, e.g., turbidity in the lower troposphere that absorbs and scatters light out of the incoming beam.

Sources of light in the sky yield corresponding horizontal illuminances that span many orders of magnitude. To the extent that total NSB is related to the total ALAN emission on the ground, it also serves as a predictor of surface illuminance, which has ecological implications. H\"{a}nel et al.~\cite{Hanel2018} tabulated literature values for conditions ranging from full daylight to naturally dark nighttime conditions, which we reproduce here in an abbreviated form in Table~\ref{light-levels}.
%
%
\begin{table}[tbph]
\begin{center}
\caption{Typical horizontal illuminances ($E_{V,H}$) and corresponding zenith luminances ($L_{V,\textrm{zenith}}$) from the literature, adapted from Table 2 in~\cite{Hanel2018}. Lighting sources are noted in column 1. Base-ten logarithms of $E_{V,H}$ and $L_{V,\textrm{zenith}}$ are given in columns 2 and 3, respectively. Where no measurement is available or a given entry has no physical meaning, an em dash (\textemdash) is shown.}\label{light-levels}%
\begin{tabular}{@{}lcc@{}}
\toprule
\textbf{Lighting} & \textbf{Log$_{10}$ $E_{V,H}$} & \textbf{Log$_{10}$ $L_{V,\textrm{zenith}}$}\\
\textbf{condition} & \textbf{(lux)} & \textbf{(cd m$^{-2}$)}\\
\midrule
Direct sunlight & 5.11 & --- \\
Overcast day & 2.00 to 3.30 & 1.51 to 2.81 \\
Extremely overlit street & 1.85 to 2.18 & 1.00 \\
End of civil twilight & 0.53 & $-0.35$ \\
Full moon in zenith & $-0.49$ & --- \\
Urban night sky (overcast) & $-1.52$ to $-0.26$ & $-2.05$ to $-0.77$ \\
Quarter moon in zenith & $-1.60$ & --- \\
End of nautical twilight & $-2.09$ & $-2.72$ \\
Urban night sky (clear) & $-2.15$ to $-1.19$ & $-2.64$ to $-1.68$ \\
Suburban night sky (overcast) & $-2.22$ to $-0.85$ & $-2.68$ to $-1.37$ \\
Milky Way (center) & --- & $-2.71$ \\
Suburban night sky (clear) & $-2.70$ to $-1.35$ & $-3.12$ to $-1.85$ \\
Rural night sky (clear) & $-3.15$ to $-2.52$ & $-3.60$ to $-3.10$ \\
Rural night sky (overcast) & $-3.15$ to $-3.05$ & $-3.60$ to $-2.57$ \\
Naturally starlit night & $-3.22$ to $-3.05$ & $-3.70$ to $-3.52$ \\
Overcast natural night & $<-3.22$ & $<-3.70$ \\
\botrule
\end{tabular}
\end{center}
\end{table}

%
%
\subsubsection{Airglow and aurorae}{\label{subsubsec:airglow-aurorae}}
In directions away from the ecliptic, these two energetic processes in the Earth's upper atmosphere are the most significant contributors to the total brightness of the natural night sky. The two mechanisms are distinguished primarily by the source of excitation: airglow derives from photoionization followed by radiative recombination, photochemical reactions, or ambient collisional excitation of atoms, whereas aurorae result from the collisional excitation of atoms by solar charged particles spiraling down the Earth's magnetic field lines. Airglow dominates at virtually all latitudes, while aurorae are most important at high magnetic latitudes. Both light sources are temporally variable in terms of their distribution on the sky, and their intensities vary on timescales ranging from seconds to hours.

In addition to the polar aurorae, an adjunct phenomenon known as a mid-latitude stable auroral arc (MSAR) yields a lower-intensity glow seen across hundreds to thousands of kilometers on the ground. It is thought to result from impacts of magnetospheric electrons accelerated downward along the Earth's magnetic field. Unlike the polar aurorae, MSAR excitation favors the $^{1}D$ transition of O~\textsc{i} and emission in the 630 nm doublet.~\cite{Borovsky2019} Another transient subauroral arc only recently identified is the Strong Thermal Emission Velocity Enhancement (STEVE), which seems to be related to ion drift in the ionosphere.~\cite{GallardoLacourt2018}

As observed and photographed, the airglow often displays a periodic structure attributable to gravity waves in the atmosphere.~\cite{Tarasick1990} These waves can be generated by phenomena ranging from volcanic eruptions to turbulence in the stratosphere induced by the action of supercell thunderstorms (Figure~\ref{airglow}b and c). Shepherd and Cho \cite{Shepherd2017} presented orbital O \textsc{i} $\lambda$577 nm observations indicating airglow enhancements up to a full order of magnitude over quiescent conditions when multiple zonal components of gravity waves come into phase at the same longitude.

The strength of the airglow contribution depends on the phase and amplitude of the solar cycle. Its time average can be approximated in S$_{10}$(vis) units according to 
\begin{equation}
B_{airglow} = 145 + 108(S-0.8)
\end{equation}

\noindent
where $S$ is the solar 10.7-cm flux in units of MJy.~\cite{Leinert1998} As $S$ is observed to vary roughly sinusoidally between 0.8 and 2.0 during the solar cycle, the quiescent airglow contribution ranges between $145-270$ S$_{10}$(vis). Excluding aerosol extinction in the lower atmosphere, natural NSB can therefore vary by a factor of nearly two both within one night and from night to night. 

NSB is also observed to vary seasonally,~\cite{Patat2008} and certain episodes seem to correlate with solar activity even near solar minimum~\cite{Grauer2021}. At least some of the seasonal variability is attributable specifically to changes in the strength of the airglow lines.~\cite{Hart2018} There are further indications of natural NSB enhancements that correlate in time at observing stations separated by thousands of kilometers.~\cite{Grauer2019} These observations highlight the importance of temporal sampling and the ongoing need to develop a deeper understanding of airglow physics. 

After accounting for line sources of radiation, there is an observed residual airglow ``continuum'' that shows no spectroscopic structure.~\cite{Broadfoot1968} It is spectrally flat over the visible wavelengths and adds no more than about 50 S$_{10}$(vis) to the brightness of the night sky.~\cite{Sternberg1972} The source of this pseudo-continuum is not entirely clear, in part because it is exceedingly difficult to fully distinguish it from zodiacal light and integrated starlight. Roach and Gordon concluded that the source of this light is a combination of ``a real upper-atmosphere phenomenon'' and sunlight multiply scattered far into the Earth's night-side atmosphere. Kenner and Ogryzlo~\cite{Kenner1984} reviewed various proposed chemical reactions involving compounds of oxygen and nitrogen, along with their own laboratory data, to explain this emission. Bates \cite{Bates1993} argued that the continuum is formed by a multitude of products resulting from collisions between metastable oxygen atoms and air molecules, ``thereby forming complexes that dissociate by allowed radiative transitions.'' The  consensus among researchers now is that the airglow continuum is composed of many hundreds of individual atomic and molecular spectral lines whose sharp features overlap to create a broad emission continuum.

\subsection{Anthropogenic skyglow}{\label{subsec:skyglow}}
Light from human-caused sources comprises the balance of NSB when added to the spatially and temporally variable contributions of natural sources. The visible manifestation of this light, commonly referred to as ``skyglow'', forms in the lower atmosphere as a result of both small- and large-particle scattering. ALAN emitted on the ground influences skyglow through its spectral power distribution, angular emission function and the total lumen output of contributing light sources.~\cite{Kocifaj2020} 

\begin{figure}[tbph]
\begin{center}
\includegraphics[width=88mm,height=35mm,angle=0]{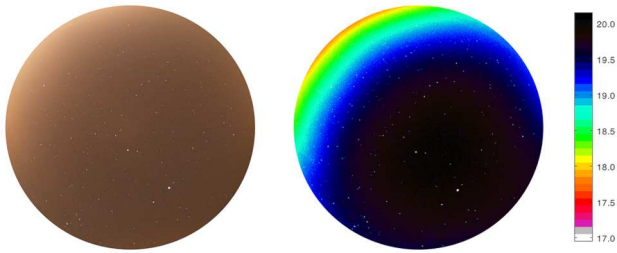}
\caption{The color and brightness of the light-polluted night sky. Uncalibrated (left) and luminance-calibrated (right) images of the sky on UT 23 May 2017 as seen from Mission San Xavier del Bac near Tucson, U.S. The zenith is at the center of both images and the horizon runs around the edge; north is at top and east at left. Warm tones in the uncalibrated image indicate the color of skyglow then dominated by high-pressure sodium lighting emissions. False colors at right correspond to luminance in units of m$_{V}$ arcsec$^{-2}$ according to the color bar at right. Images by the author.\label{SanXavier-AllSky-170522}}
\end{center}
\end{figure}

Skyglow is observed to vary in brightness relative to the assumed zenith brightness of an unpolluted night sky by up to a factor of about 6,000,~\cite{Jechow2020} at which point only a handful of the brightest stars are visible to the unaided human eye. In most urban contexts, skyglow dominates the light of the night sky (Figure~\ref{SanXavier-AllSky-170522}).

The presence of ice and snow on the ground intensifies skyglow due to their high reflectivities, enhancing upward-directed emissions from cities; models of skyglow formation over cities show an almost linear relationship between ground reflectance and artificial NSB.~\cite{Aube2015} Measurements of the effect show an up to three-fold increase in NSB in cities due to snow cover on the ground,~\cite{Falchi2010} and snow cover further amplifies skyglow itself due to reflections of the sky from the ground.~\cite{Jechow2019,Wallner2019} Skyglow is also sensitive to the presence of very fine particles in the air, which may be increased by certain kinds of air pollution.~\cite{Liu2020,Kocifaj2021} Cloudy nights make the problem even worse;~\cite{Sciezor2020} overcast conditions over cities increase horizontal illuminance at ground level by a factor of up to ten.~\cite{Kyba2011} On the other hand, the comparative absence of ALAN in rural places means that cloud cover tends to darken the nighttime sky and landscape. On an overcast night far from sources of anthropogenic light, the measured NSB can be up to about a factor of six lower than a clear night at the same location.~\cite{Jechow2019b}

Skyglow is now well understood as a matter of radiative transfer, and models have steadily become more sophisticated and better representative of real-world conditions. The total light output of a city is the strongest predictor of NSB in the urban environment, and of the brightness of `light domes' over cities as seen from remote locations.~\cite{Luginbuhl2009} Modelers have also attempted to infer the so-called city emission function (CEF) as a means of describing the distribution of the anthropogenic light over a city causing skyglow.~\cite{Kocifaj2015,Kocifaj2018,Kocifaj2019} Understanding the CEF is crucial for predicting the appearance of skyglow both within and outside of the city; for example, light rays emitted at very shallow upward angles yield the greatest impact on NSB as seen at distances from dozens to hundreds of kilometers.~\cite{LuginbuhlBoleyDavis2014} Shadowing of city light emissions by topographic features is also known to influence the CEF.~\cite{Cinzano2001,Aube2015}

\subsection{The scattering of light in the Earth's atmosphere}\label{subsec:scattering}
The behavior of light during its flight through the atmosphere is governed by the frequency-dependent radiative transfer equation:
\begin{equation}
\frac{dI_{\nu}}{d\tau_\nu} = -I_{\nu} + S_{\nu}
\end{equation}
where $I_{\nu}$ is the light intensity, $\tau_{\nu}$ is the optical depth, and $S_{\nu}$ is the so-called ``source function'' defined by the ratio of emission and absorption coefficients.~\cite{RybickiLightman1985} In the simple case of a plane-parallel atmosphere, or otherwise in conditions where the atmosphere is horizontally homogeneous, it has the general solution.~\cite{Schafer1980}
\begin{equation}
I_{\nu}(\tau_{\nu}) = I_{\nu}(0) e^{{-\tau_{\nu}}} + \int_{0}^{\tau_{\nu}} S_{\nu} e^{-({\tau_{\nu}-{\tau^{\prime}_{\nu}}})} S_{\nu}(\tau^{\prime}_{\nu}) d\tau^{\prime}_{\nu}.
\end{equation}

This formula governs the frequency-dependent intensity of light as it traverses a medium with optical depth $\tau_{\nu}$. In short, it holds that the change in intensity during the traverse is the sum of light added to the beam less the sum of light removed from the beam. Light can be added through direct emission from the medium and removed via absorption and scattering. 

The remainder of this section focuses on the scattering of light by atmospheric constituents, as this process dominates the radiative transfer process in determining NSB. We assume local thermodynamic equilibrium, which is a reasonable approximation for clear sky/air conditions, and for the moment we neglect both complete absorption of light and direct emission from the atmosphere through, e.g., airglow.

Two modes of scattering control the behavior of light as it transits the atmosphere: Rayleigh scattering and Mie scattering. These modes depend largely on the size of the scattering particles, which are composed of two principal atmospheric constituents: molecules and `aerosols' (a suspension of fine solid particles or liquid droplets in air). Rayleigh scattering is important when the wavelengths of light involved are much smaller than the size of the scattering particles. The scattering strength is strongly dependent on wavelength ($I \propto I_{0}\lambda^{-4}$). Mie scattering applies exclusively to homogeneous, spherical particles and shows almost no scattering strength dependence on wavelength; in comparison, the particles in the Earth's atmosphere are distinctly inhomogeneous and irregular in shape. To the extent that atmospheric particles are comparable in size to the wavelength of light, rather than much smaller or much larger, circumstances are reasonably approximated by Mie scattering. The distinction between Rayleigh and Mie scattering is obvious in everyday experience: the blue color of the daytime sky results from strong Rayleigh scattering of short-wavelength visible light by diatomic nitrogen molecules comprising the majority of the lower atmosphere, while clouds are white or gray depending on whether they reflect or transmit (attenuated) sunlight. 

Short-wavelength light is efficiently Rayleigh-scattered even along short optical paths, yielding blue-rich spectra near light sources and progressively `de-blued' spectra at large distances. Rayleigh scattering dominates NSB in both cases. Mie scattering becomes an important influence in and near cities, especially where particulate pollution from vehicle exhaust and industrial activity are common.~\cite{LuginbuhlBoleyDavis2014,Aube2015} Furthermore, multiple-order scattering is often important in real-world situations. Sophisticated radiative transfer codes account for this in skyglow models, although they tend to become processor-intensive as the number of scattering orders increases. 

Some light exits the atmosphere completely, whether directly or after one or more scattering events, and can be detected from orbit. This forms the basis of remote sensing of upward radiance, which is discussed in Section~\ref{subsubsec:remote-sensing}.

%
%
\section{Existing sensing and monitoring capabilities}\label{sec:sensing-monitoring}

The measurement of NSB from ground-based platforms has been recently reviewed by H\"{a}nel et al.~\cite{Hanel2018} We summarize the relevant points here.

\subsection{Sensing}\label{subsec:sensing}

There are two basic approaches to measure and monitor NSB: look upward from the ground or look down from orbit. The former mode involves direct sensing of NSB, while the latter mode predicts NSB seen from the ground by sensing upward-directed radiance and applying a model of how light propagates through the atmosphere. Raw ground-based measurements are model-independent but typically limited geographically and temporally. We focus here largely on the ground-based approach, but briefly comment on new capabilities for remotely sensing NSB in Section~\ref{subsubsec:remote-sensing}.

Direct measurements of NSB from the ground involve sensors that integrate the flux of light through a known solid angle, within some wavelength range, and over some length of time. These divide into two types: single-channel devices and multichannel devices. 

\subsubsection{Single-channel devices}\label{subsubsec:single-channel}
Single-channel devices are patterned on photoelectric photometers used by astronomers for almost a century. These devices, such as the popular Sky Quality Meter (SQM;~\citealp{Cinzano2005,Cinzano2007}), rely on simple and well-understood physics, require little electric current to operate, and are usually small enough to be easily portable. They typically employ light-to-frequency (LTF) converters whose output is a signal pulse stream, the frequency of which is linearly proportional to received light intensity. Their light response is determined in the laboratory, with on-board lookup tables relating measured frequency to light intensity tied to calibrated light sources. Since the response of LTF converters is also sensitive to ambient operating temperature, sensing of the air temperature is required to properly correct the measured frequency. This is usually done on board the measurement device.

Most commercially available devices have their own photometric passbands modeled on Johnson-Cousins $V$.~\cite{Bessell1990} Researchers have experimented with other filters, but $V$ was chosen to match the bulk of existing literature data and the human visual response to light under photopic conditions. Infrared blocking filters are often used in combination with the quantum efficiency profile of the semiconductor material of the LTF to achieve the desired effective passband. Optics may be used to constrain the opening angle defining the device's angular field of view. Although single-channel device measurements indicate only the brightness of the night sky averaged across a fairly large acceptance angle, some authors report creating crude two-dimensional maps of NSB by interpolating spot measurements from these devices.~\cite{Zamorano2014}

Single-channel devices have a number of advantages, including ease of use; portability; a physically simple sensing mechanism; temperature compensation; good repeatability; rapid capture and display of data; and a relatively long historical record of use. However, there are certain drawbacks to these devices. In order to sense a sufficient amount of light to yield a measurable signal, they must integrate it over a relatively large solid angle. They offer little meaningful spatial resolution in most applications, making them generally unsuitable for monitoring the behavior of light domes near the horizon. Lastly, there are differences among commercially available devices in terms of photometric passbands that complicate comparison of results among different device types.

\subsubsection{Multichannel devices}\label{subsubsec:multichannel}

\begin{figure}[tbph]
\begin{center}
\includegraphics[width=86mm,height=68mm,angle=0]{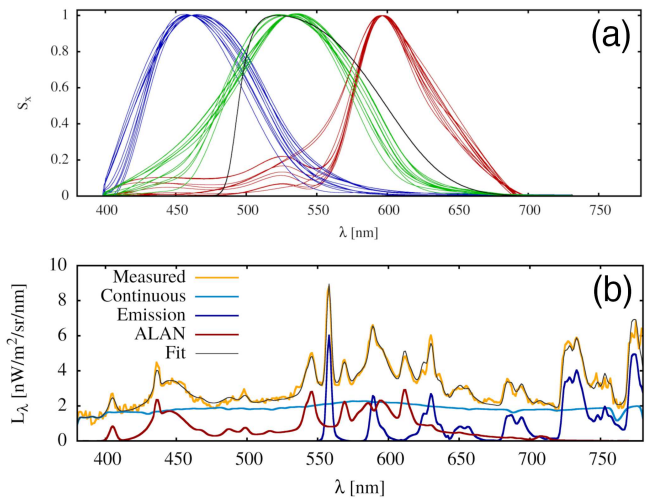}
\caption{Typical broadband digital imaging passbands and night sky spectra. (a) Representative spectral sensitivity curves of some commercial digital cameras (blue, green and red lines corresponding to broad RGB bands) and the astronomical \emph{V}-band response (black line). Figure 2 in~\cite{Kollath2020}. (b) The night sky spectrum over Zselic International Dark Sky Park, Hungary (yellow) and its decomposition and fit (other colors). See main text for a description of the fit components. Figure 1 in~\cite{Kollath2020}.\label{Kollath-et-al-2020}}
\end{center}
\end{figure}

Multichannel detectors consist of arrays of light-sensitive elements whose output is multiplexed through one or more signal amplifiers. The ideal example is an imaging spectroradiometer, which provides a complete set of information about the wavelength-dependent brightness of the night sky in any given direction. However, the current generation of such devices is too slow for capturing time-resolved NSB data, and they tend to be prohibitively expensive. One more often encounters cameras capturing two-dimensional images, particularly commercial digital single-lens reflex (DSLR) cameras and mirrorless interchangeable lens cameras (MILC); see, e.g., ~\citealp{Falchi2010}. Some are operated with photometric filters to yield a particular effective passband, while others use Bayer filter mosaics to capture native (pseudo-)true-color images through the combination of broadband red-, green- and blue-filtered data. (Figure~\ref{Kollath-et-al-2020}a) 

The main advantage these cameras have over single-channel devices is the ability to produce two-dimensional images with some amount of both angular and spectral resolution. They are often paired with very wide-angle lenses to capture views with solid angles as large as 2$\pi$ steradians (180$^{\circ}$) in a single exposure,~\cite{Aceituno2011} while others build up multiple-image mosaics with angular offsets between exposures so that the results can later be ``stitched'' together in software.~\cite{Duriscoe2007} As a result, these devices provide significantly more spatial information about the distribution of NSB than do single-channel devices. 

Depending on the pixel scale of the detector, star images may be sufficiently sampled that flux calibration can be performed using spectrophotometric standard stars; other imaging systems make use of lab calibrations from reference light sources and employ integrating spheres for illumination of the camera and lens. Spatial distortion information for particular lens and camera combinations can be used to correct lens aberrations after the fact in software.~\cite{Mohar2015,Kollath2017}

Multichannel devices have their own drawbacks. Due to sensor size and pixel scale, they generally have limited angular resolution. When imagers are used with fisheye lenses to capture all-sky data in single exposures, significant angular distortions are induced near the horizon. Their multi-spectral functionality is usually limited to a few broad passbands. And, lastly, there is as yet no standard, SI-traceable reporting unit for NSB measurements. This issue is discussed further in Section~\ref{sec:units}. 

\subsubsection{Color considerations}\label{subsubsec:color}
An concern adjunct to characterizing NSB is the spectral power distribution (SPD) of sky light. As the preceding discussion suggests, the sensed NSB is the result of integrating, with respect to wavelength, the convolution of the SPD of the night sky with the spectral bandpass of the measuring device. The SPD of the night sky is a complex function of the various physical processes from which it results (see Secton~\ref{sec:nsb-components}); it is further modulated by wavelength-dependent scattering during the transit of night sky light through the Earth's atmosphere. Measurements of NSB in both radiometric and photometric units are therefore strongly dependent on the night sky's spectrum.~\cite{Bara2020} Because most devices used to sense NSB have relatively large spectral bandpasses, the responses of those instruments interact with the night sky SPD in complex ways and call for careful consideration when interpreting measurements.~\cite{Sanchez2017}

Some authors report the use of metrics such as the correlated color temperature (CCT) of the night sky as a means of characterizing its spectral qualities.~\cite{Jechow2019b,Jechow2020} While CCT relates to the spectra of thermal sources, its utility is diminished as the SPDs of sources become increasingly non-thermal. Since many NSB components, such as airglow and aurorae, have decidedly non-thermal SPDs, the use of CCT alone is unlikely to give reliable color information about the night sky.

\subsubsection{Data modeling}\label{subsubsec:data-modeling}

Modeling of NSB observations can assist with their analysis and interpretation. For example, Duriscoe~\cite{Duriscoe2013} reported successfully recovering the anthropogenic component of NSB from mosaicked all-sky image data by subtracting 2-D models of natural sources of light. To the extent that construction and application of such models can be automated, they hold the promise of rapidly disentangling natural sources of light in the night sky from artificial sources for the purposes of modeling the angular and temporal evolution of skyglow.

For spectrally resolved measurements, it is possible to model the natural components of NSB in wavelength space to subtract and remove them, leaving behind only the spectrum of artificial light sources. Figure~\ref{Kollath-et-al-2020}b shows the decomposition of a night sky spectrum containing both natural light sources and anthropogenic skyglow. The “Continuous” component (light blue) is the sum of contributions from natural light sources other than airglow line emission. “Emission” (dark blue) consists of modeled line airglow contributions. “ALAN” (red) is the sum of several common lamp spectra. The sum of all components is the “Fit” line (gray). From this decomposition it was determined that the `continuous' component of the natural sky (zodiacal light, scattered starlight and airglow pseudo-continuum) is nearly constant at all visible wavelengths and has a spectral radiance of $\sim2$ nW m$^{-2}$ sr$^{-1}$ nm$^{-1}$.~\cite{Kollath2020}

There are a handful of additional approaches to the modeling NSB that add other inputs to the direct sensing of light. For example, Koll\'{a}th and Koll\'{a}th~\cite{KollathKollath2020} used raw backscatter data from a laser ceilometer to provide inputs to Monte Carlo simulations of sky radiances measured simultaneously from the ground using calibrated cameras. The authors applied this technique to infer the vertical structure of the radiance distribution of the night sky.

\subsubsection{Remote sensing of NSB}\label{subsubsec:remote-sensing}

NSB is now routinely measured by remotely sensing upward-directed radiance using a variety of platforms, including Earth-orbiting satellites,~\cite{Levin2020} the International Space Station~\cite{Sanchez2019}, airplanes,~\cite{Barducci2003,Kuechly2012} drones~\cite{Li2020} and balloons.~\cite{Walczak2021,Bettanini2022} The use of remote sensing to infer NSB in this manner offers a number of attractive qualities. Chief among these is the ability to collect information about NSB from essentially anywhere on Earth, which decouples NSB measurement and monitoring from the deployment of ground-based sensors. Falchi and coworkers provided such a global data product most recently in 2016.~\cite{Falchi2016} They calibrated the radiance-NSB relationship using many thousands of ground-based NSB measurements, but their predictions are sometimes inaccurate. This may be the result of models assuming a flat Earth, and which therefore do not take into account the screening effect of regional topography, or due to the fact that locally variable atmospheric turbidity can induce time-dependent scattering effects. In particular, at astronomical observatories, which tend to be located in comparatively dry, high-altitude sites, the aerosol content is probably overestimated as therefore also is the computed scattering. Yet this map remains our only truly global view of light pollution.

Diffuse light seen around cities in remote sensing imagery from Earth orbit was long thought to result from a combination of sensing artifacts and low spatial resolution,~\cite{LImhoff1997,Li2017} but it is now recognized as a real signal corresponding to light scattered in the atmosphere. Kocifaj and Bar\'{a}~\cite{KocifajBara2020} showed that certain aerosol properties, such as the particle size number distribution, can be successfully retrieved from orbital radiometry of the angular radiance distribution of the scattered light near cities. S\'{a}nchez de Miguel et al.~\cite{Sanchez2020} recently found a strong correlation between the zenith NSB measured on the ground and orbital radiance measurements at both low and high resolution. They suggested that creating accurate, regional or even global NSB maps based on radiance measurements from the newest generation of orbital radiometers should be possible.

However, there are other problems with existing satellite remote sensing platforms. For example, the Day-Night Band of the Visible Infrared Imaging Radiometer Suite (VIIRS-DNB), deployed on board the \emph{Suomi NPP} and \emph{NOAA-20} satellites, has no spectral sensitivity shortward of 500 nm. The instrument is therefore effectively blind to the strong peak in white LED light emissions near 450 nm. This limits what can be reliably inferred concerning short-wavelength light sources within the data set.~\cite{Bara2019}

\subsection{Monitoring}\label{subsec:monitoring}
In context of this Review, ``monitoring'' of NSB refers to its repeated measurement to look for trends on timescales ranging from minutes to years. Monitors, like sensing devices, fall into two general categories: those that function autonomously, and those whose operation requires human attendants.

Autonomous monitors are sensing devices fitted into weatherproof housings with their own electric power supplies and, optionally, network connections. Some of them save their measurements to on-board memory, while others relay them to another location for storage via a local network or the Internet. At present, autonomous monitors tend to be single-channel devices with few requirements for field calibration. These monitors are subject to regular insolation during the daytime, which appears to contribute to photometric zeropoint drift, possibly through the deterioration of optical and/or electronic components due to solar ultraviolet light irradiation.~\cite{BaraSQM2020} Some authors report attempts to calibrate these long-term secular trends using luminance sources like the twilight sky.~\cite{Puschnig2020}. 

Attempts to construct autonomous all-sky imagers have tended to leverage existing facilities marketed to amateur astronomers as cloud sensors; other, purpose-built devices, such as the ASTMON system~\cite{Aceituno2011}, are intended as fully robotic instruments whose data acquisition and reduction are automated and which function as permanent monitors.

Attended monitors may function automatically, but they require a human operator for setup and maintenance. This is usually because the monitoring device is not permanently installed and lacks equipment to make it durable in the natural environment. The operator may also direct details of the data collection protocol such as manually switching slides in a rotating filter wheel. An example of this is the Road Runner system, in which a single-channel sensing device is mounted to the roof of an automobile and collects NSB data continuously while the vehicle is driven.~\cite{RosaInfantes2011} Another example is the U.S. National Park Service Night Sky Team (NPS NST) imaging method.~\cite{Duriscoe2007} Its camera, situated on an automatic `go-to' mount, executes an imaging program automatically, but it must be transported to each measurement site and set up by NPS NST staff. There is also considerable human effort required to reduce, analyze, and report the resulting data.

Monitoring entails the concerns of data handling, transmission and storage, as well as reduction and analysis. Some autonomous monitors log NSB data to on-board storage media, which must be periodically retrieved and offloaded. Other systems, such as the Telescope Encoder and Sky Sensor-WiFi (TESS-W),~\cite{Zamorano2017} make use of wireless networking and transmission of measurements to a central storage location via the Internet, leaving them vulnerable to network interruptions. There are also concerns about data reporting formats, although some effort has been put into designing and promoting a standard protocol for recording NSB data.~\cite{Kyba2012}

\subsubsection{Temporal sampling frequency}\label{subsubsec:temporal-sampling}

Other monitoring considerations involve the frequency of data collection, both in the temporal and spatial senses. Given the timescales on which the natural NSB varies, sampling frequency is important so as to fully understand the brightness range of the natural nighttime environment; the same applies to skyglow, which tends to vary in slower and more predictable ways. The presence of skyglow can `stabilize' NSB if it significantly exceeds the radiance of  natural sources of light in the night sky, as in many bigger cities. In such cases, only weak apparent variations exist from night to night. NSB monitors therefore typically perform best in urban environments while potentially giving ambiguous information in naturally dark locations.

Various approaches to visualizing NSB time-series data are suggested in the literature. Perhaps the most common method is the NSB densitogram, commonly referred to as a `jellyfish plot'. In this representation, the NSB is plotted against the local time, and each pixel is color-coded to represent the number of observations in a time series that fall into that particular (time, NSB) bin. It is a convenient way to both compress a lengthy time series into a single plot as well as to quickly discern between typical and atypical NSB conditions.

\begin{figure}[tbph]
\begin{center}
\includegraphics[width=88mm,height=134mm,angle=0]{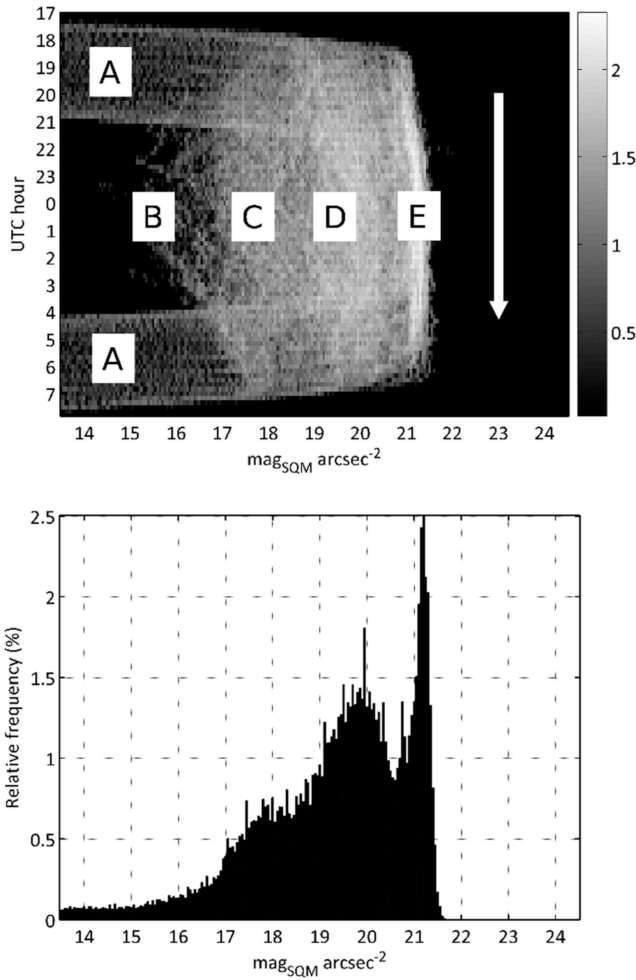}
\caption{NSB histogram made by integrating a time series of measurements obtained over one year at Paramos, Spain. The upper plot demonstrates the method schematically, with the result shown in the lower plot. ‘Arms’ on either side of the plot (“A”) represent the rapid sky brightness change during twilight. The brightest nights (“B”) are those during which moonlight directly illuminates the detector. “C” represents nights during which clouds scatter moonlight and reflect city lights. Nights “D” correspond to moonless nights when clouds amplify city skyglow. The darkest nights (“E”) correspond to clear conditions with no moonlight contribution. Figure 2 in \cite{Bara2019}.\label{Bara-Lima-Zamorano-2019-fig2}}
\end{center}
\end{figure}
This kind of data visualization helps inform efforts to characterize night sky quality at a given location and follow its evolution in time. For example, Bar\'{a} et al.~\cite{Bara2019} suggest that a well-sampled jellyfish plot can be used to extract meaningful sky quality metrics. In Figure~\ref{Bara-Lima-Zamorano-2019-fig2}, a jellyfish plot is collapsed to a 2-D distribution in frequency versus zenith NSB. Peaks corresponding to the various brightness regimes are evident in the result. From this, the authors conclude that no single value of the NSB fully represents the variety of conditions at any particular observing site.

Some limited efforts have been made to apply, e.g., Fourier analysis techniques to time-domain measurements of NSB. For instance, Puschnig, Wallner and Posch~\cite{Puschnig2019} used fast Fourier transform frequency analysis of nightly mean NSB measurements made using a network of Sky Quality Meters in Austria. From this analysis they concluded that the circalunar periodicity of NSB, of biological importance to a number of nocturnal species, essentially disappears for maintained zenith brightnesses higher than about 16.5 mag arcsec$^{-2}$ ($\sim32$ mcd m$^{-2}$).

Bar\'{a} et al.~\cite{Bara2019} further considered whether the NSB sampling rate on a timescale of minutes influences average sky quality indicators using measurement collected in long (e.g., yearly) time periods, concluding that it does not. Resampling a series of zenith brightnesses obtained with Sky Quality Meters in one-minute readings to sampling intervals of five and ten minutes, they found that the the maximum absolute difference of the full width at half-maximum (FWHM) of the darkest peak in a histogram of time-series NSB values was $<0.0009$ mag arcsec$^{-2}$ for a five-minute sampling interval and $<0.0017$ mag arcsec$^{-2}$ for a ten-minute sampling interval. These values are well below the measured precision of the SQM ($\lesssim5$\%). 

However, the question of which temporal NSB sampling frequencies are sufficient to yield a sense of the typical night sky quality at a given location is not well formed because there is yet no general agreement as to what we mean by `typical'. If this were clearly and definitively decided, a simple analysis would easily reveal the optimal sampling parameters to yield the desired metric. An example of how this approach may be applied to NSB data is discussed in Section~\ref{subsubsec:nsb-histograms}. 

\subsubsection{Spatial sampling frequency}\label{subsubsec:spatial-sampling}

Characterizing the typical NSB across a large geographic area demands consideration of the proper spatial sampling frequency in order to ensure uniform results, especially with respect to acceptable measurement uncertainties. To date there is one published study on this subject by Bar\'{a}~\cite{Bara2017}, based on data from Falchi et al.~\cite{Falchi2016} Bar\'{a} found that a useful rule of thumb is that one measurement per square kilometer is sufficient to constrain the zenith NSB at any point in a sampled region to a precision of $\pm0.1$ $V$ mag arcsec$^{-2}$ rms. However, the author notes that ``exact reconstruction of the zenithal night sky brightness maps from samples taken at the Nyquist rate seems to be considerably more demanding.''

%
%
\section{Measurement units}\label{sec:units}

NSB measurements found in the literature are reported in several different, and sometimes confusing, units. Although one occasionally finds illuminances reported in SI units like microlux, the majority of measurements are surface brightness terms. As a further complication, measurements can be either radiometric or photometric depending on whether they refer broadly to the entire visible spectrum or instead are weighted by the spectral response of the human eye, respectively. Some units characterizing NSB in surface brightness terms are as follows:
\begin{itemize}
\item \textbf{Candela per square meter} (cd m$^{-2}$), a linear, SI unit informally called the ``nit''. The unit is based on the SI units of luminous intensity (candela) and area (meter). The CGS equivalent is the \bf stilb. \rm 1 stilb = $1.04\times10^{2}$ cd m$^{-2}$.
\item \textbf{Lambert} (L), a linear, non-SI unit defined as $\pi^{-1}$ cd cm$^{-2}$ ($\approx$ 3183 cd m$^{-2}$).
\item \textbf{S$_{10}$(vis)}, a linear, non-SI unit defined as the surface brightness of a $m_{V}=+10$ star whose light is distributed over one square degree. 1 S$_{10}$(vis) $\approx$ $1.04\times10^{-6}$ cd m$^{-2}$.)
\item \textbf{Magnitude per square arcsecond} (mag arcsec$^{-2}$, or mpsa), a logarithmic, non-SI unit defined such that if an area on the sky contained only exactly one magnitude $N$ star in each square arcsecond, the sky brightness would be $N$ mag arcsec$^{-2}$.
\item \textbf{Night Sky Unit} (NSU), a linear, non-SI unit introduced in Section 5.2 as the average zenith NSB away from the ecliptic assuming quiescent airglow conditions and the absence of skyglow ($\sim0.2$ mcd m$^{-2}$ in the $V$ band). It is sometimes called a ``Natural Sky Unit'' or a ``sky''.
\end{itemize}

Of these, the magnitude per square arcsecond is most often encountered, being the native reporting unit of, among other devices, the popular Sky Quality Meter. Transformations between magnitudes per square arcsecond and SI luminance units have been derived so that astronomical brightnesses in, e.g., $V$ magnitudes can be approximately transformed to photometric values. Noting that the relationship between these quantities depends on the spectral power distribution of the source, transformation equations have been derived for scotopic~\cite{Bara2020} and mesopic~\cite{Fryc2021} viewing conditions and calibrated using zero-point luminances determined from a variety of night sky spectra.

Koll\'{a}th et al.~\cite{Kollath2020} lately discussed the problem of different effective photometric passbands among both single-channel and multichannel devices used to measure NSB, as well as the lack of standardized, SI-traceable reporting units. Since the range of the band-averaged spectral radiance of a device is independent of the selected passband for spectrally flat or constant sky radiance, the measured band-averaged spectral radiance is of the same order and takes the SI unit of W m$^{-2}$ sr$^{-1}$ m$^{-1}$. Given typical NSB values and the wavelengths of light involved, a more natural unit is nW m$^{-2}$ sr$^{-1}$ nm$^{-1}$, which the authors propose as one Dark Sky Unit (DSU). In this unit, the band-averaged radiance of the natural night sky under clear-air and quiescent-airglow conditions is approximately 1--2, while a cloudy sky yields a value of about 1. These numbers are applicable for any passband defined in the visible spectrum. On nights when airglow is particularly active and its spectrum is dominated by line emission rather than the pseudo-continuum, it can increase the band-averaged radiance at the zenith by almost a factor of two as compared to nights when airglow is relatively inactive.

%
%
\section{Classifying and ranking night sky quality}\label{sec:ranking}
Measurement and monitoring of NSB are usually conducted to meet one or more objectives. These may involve gathering a baseline of nighttime conditions to initially characterize the quality of the night sky before monitoring begins, with an eye toward assessing the impact of skyglow on an area. This supports site selection for new ground-based astronomical observatories. Long-term monitoring helps site managers identify potential threats to night sky quality and assess the efficacy of various mitigations.

While NSB can be quantified, there is no fully objective quality determination or ranking system for the night sky in part because the experience is distinctly human-focused. Any attempt to compare or rank sky quality must admit the inability to completely specify what for many people is an emotional, psychological and sensory experience. Furthermore, we now know that the natural night sky varies considerably in brightness on timescales ranging from minutes (aurorae) to years (atmospheric extinction from events like volcanic eruptions). As a result, relative measurements that look for trends in NSB at a single location are often more reliable than meta-analyses that aim to quantitatively compare two or more locations.

The aerosol content of the lower atmosphere is an important factor controlling the perceived quality of the night sky for two reasons. One is that aerosols are responsible for the direct attenuation of light from astronomical objects, making them appear to be less bright than they would be above the Earth's atmosphere. The other reason is that aerosols scatter light -- whether natural or artificial -- and raise the level of the sky background. These effects jointly act to reduce the contrast between astronomical objects and the background, making them difficult to see.~\cite{Crumey2014} In cities, the scattered light of skyglow overwhelmingly dominates attenuation of starlight as the cause of this contrast reduction, but in rural areas direct attenuation competes with scattering.

The importance of aerosols in otherwise naturally dark places may well explain measurements of NSB apparently falling below the naturally imposed `floor' discussed in Section~\ref{subsec:natural-light}. But with these unnaturally dark skies should come a diminished ability to see faint stars whose dim light is substantially or completely extinguished during its passage through the atmosphere. This suggests that night sky quality is ultimately determined by some combination of objective sky brightness measurements and subjective impressions of the visibility of faint astronomical objects.

\begin{figure}[tbph]
\begin{center}
\includegraphics[width=88mm,height=118mm,angle=0]{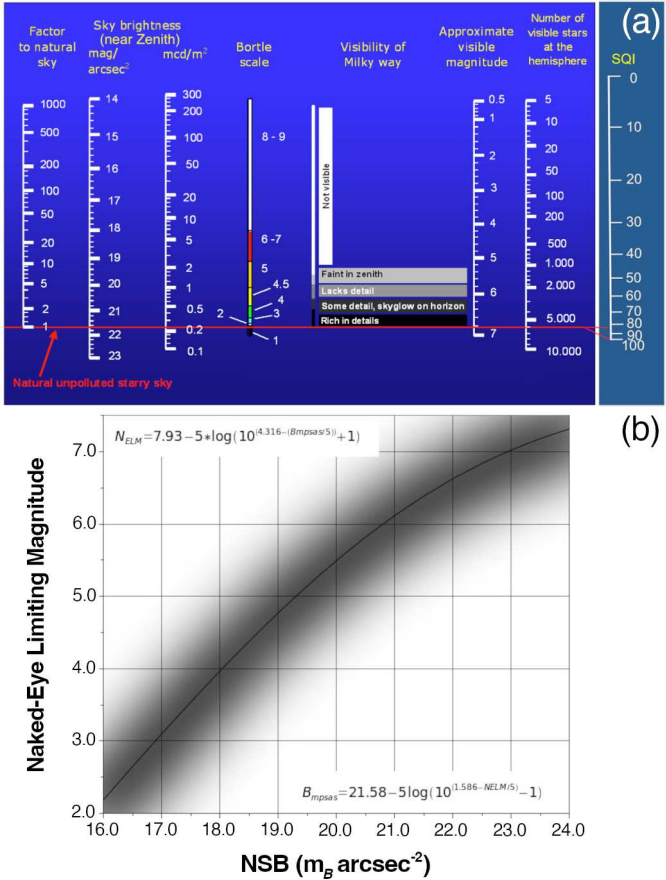}
\caption{A comparison of subjective and objective night sky quality metrics. (a) Various metrics indicating night sky quality combined into a nomogram, a type of figure in which any horizontal line displays the same quantity on multiple scales. In addition to objective measures of NSB, several subjective scales (Bortle, Milky Way visibility, and Sky Quality Index) are shown. Nomogram by Henk Spoelstra, used with kind permission. (b) An empirical calibration between naked-eye limiting magnitude and approximate Johnson-Cousins $B$-band NSB. Approximate conversion formulae to and from each variable are indicated on the plot. Data and plot courtesy of Anthony Tekatch (Unihedron).\label{visibility}}
\end{center}
\end{figure}

Both subjective (observer-dependent) and objective (device-dependent) quality metrics have been proposed and are discussed below. Several of these metrics are inter-compared in Figure~\ref{visibility}a.

%
%
\subsection{Subjective metrics}\label{subsec:subjective}

\subsubsection{Naked-Eye Limiting Magnitude}\label{subsubsec:nelm}

Subjective metrics tend to rely on the human visual system for sensing NSB. Some approaches use estimates of the visual or naked-eye limiting magnitude (NELM) as an indicator of NSB, given empirical relationships between the two (Figure~\ref{visibility}b). NELM estimates are the basis for citizen science efforts such as Globe At Night, whose data have been shown in aggregate to correctly approximate the NSB as measured through direct sensing.~\cite{Kyba2013} 

\subsubsection{Bortle Scale}\label{subsubsec:bortle}
Other subjective metrics are more impressionistic, such as the Bortle Scale,~\cite{Bortle2001} which ranks night sky quality on a scale ranging from one to nine. While the Bortle Scale is intuitive, anecdotally some users report sufficient ambiguity in the descriptions of the Bortle classes that estimates are often not reliable to better than one full step in the scheme of classes. 

The apparent brightness and degree of visual structure in naked-eye observations of the Milky Way is sometimes suggested as an alternative indicator of night sky quality. Crumey \cite{Crumey2014} argued that the fundamental visibility of the Milky Way is an indicator of what we would call a ``dark'' sky, but notes the useful limits of this idea.

%
%
\subsection{Objective metrics}\label{subsec:objective}

\subsubsection{Anthropogenic Light Ratio}\label{subsubsec:ALR}
The U.S. National Park Service (NPS) has proposed the Anthropogenic Light Ratio (ALR) as the most basic means of representing the relative amount of skyglow visible from a given site.~\cite{MooreTurinaWhite2013} It is the ratio of anthropogenic to natural NSB averaged over the entire sky. The latter quantity is usually taken as the ``night sky unit'' described here in Section~\ref{subsec:natural-light}. For purposes of computing ALR, NPS assumes an unpolluted night sky to have a zenith luminance of 78 nL ($\sim0.25$ mcd m$^{-2}$, or 21.79 mag arcsec$^{-2}$). ALR is linear and unitless, and it can be equivalently expressed as a percentage. As a consequence, comparisons between sites in terms of ALR are easily made.~\cite{MooreTurinaWhite2013} 

ALR is in fairly wide use as a NSB metric. For example, Falchi et al.~\cite{Falchi2016} used ALR as the basis for the maps presented in the New World Atlas of Artificial Night Sky Brightness. Duriscoe et al.~\cite{Duriscoe2018} presented a method for estimating ALR with high confidence over large regions using cloud-free, composite satellite images and a simplified spatial model. However, as an all-sky average, ALR fails to adequately characterize the distribution of light near the horizon where the `light domes' of cities appear. A potential adaptation of ALR that would make it more robust is to specify it as a function of altitude and azimuth. 

\subsubsection{Sky Quality Index}\label{subsubsec:SQI}
NPS has also devised and promoted the Sky Quality Index (SQI), a metric derived from the distribution of sky luminance values in all-sky image data modeled using the method of Duriscoe~\cite{Duriscoe2013}, described here in Section~\ref{subsubsec:data-modeling}, to remove the light of the natural sky and leave only the anthropogenic contribution. SQI can take any value from 0 to 100, where 100 is defined as a night sky entirely devoid of skyglow.

\subsubsection{Illuminance metrics}\label{subsubsec:illuminance-metrics}
Duriscoe~\cite{Duriscoe2013} also proposed a number of quantities derived from calibrated all-sky imagery that could be used to objectively characterize NSB at a given site. These include the average all-sky luminance (both total observed and that from skyglow alone) and the maximum horizontal and vertical illuminances. Patat \cite{Patat2003} suggested characterizing a night sky by sampling the darkest area of the sky relative to the distance between that point and the ecliptic and the galactic equator in order to minimize contributions from the zodiacal light and diffuse galactic light, respectively. Duriscoe criticized this approach because it does not account for the influence of the (spatially averaged) airglow, which increases with zenith distance by a factor of five between the zenith and the horizon.

\subsubsection{NSB histograms}\label{subsubsec:nsb-histograms}
Bertolo et al.~\cite{Bertolo2019} recently analyzed NSB data obtained by the Veneto Sky Quality Meter Network, comparing the night sky quality at seven sites in northeastern Italy. They characterized the distribution of time-series SQM measurements by the FWHM of the darkest peak of the histogram. Bar\'{a}, Lima and Zamorano~\cite{Bara2019} elaborated on this idea, proposing as a metric $m_{\textrm{\tiny{FWHM}}}$, which is the average value of the readings within the FWHM of the darkest peak of the histogram under No-Sun No-Moon (NSNM) conditions. They contrasted $m_{\textrm{\tiny{FWHM}}}$ against $m_{1/3}$, defined as the average of the upper tertile of NSB values obtained under the same NSNM conditions. While $m_{1/3}$ is advantageous in bright, generally urban contexts and suitable for long-term site monitoring, $m_{\textrm{\tiny{FWHM}}}$ is superior to $m_{1/3}$ in naturally dark locations. In the latter case, $m_{1/3}$ can be strongly influenced by very dark NSB measurements due to effects such as overcast or foggy conditions and snow cover obscuring the detector field of view.

\subsubsection{Sky brightness percentiles}\label{subsubsec:percentiles}
Hung~\cite{Hung2021} examined over 1,500 NPS Night Sky Team imaging datasets collected along with Sky Quality Meter measurements and visual observations from hundreds of U.S. national parks and monuments over nearly two decades, finding strong correlations between various commonly used night sky quality metrics. She performed a principal component analysis on 53 metrics derived from 1,391 complete datasets and concluded that only five principal components are required to explain 99\% of variations among the metrics. Hung concluded that zenith brightness and five brightness percentiles (50, 95, 99, 99.995, and 99.999) represent a minimum set that provides non-redundant characteristics of night sky quality.

\backmatter
\bmhead{Corresponding Author}
Correspondence and requests for materials should be addressed to John Barentine.
\bmhead{Acknowledgments}
Many individuals contributed positively to the content and clarity of this manuscript with their informal reviews and comments. In particular, we wish to thank Salvador Bar\'{a} (Universidade de Santiago de Compostela, Spain); Gilbert Esquerdo (Fred Lawrence Whipple Observatory, U.S.); Li-Wei Hung (U.S. National Park Service Natural Sounds and Night Skies Division, U.S.); Zolt\'{a}n Koll\'{a}th (Konkoly Observatory, Hungary); and Ken Walczak (Adler Planetarium, U.S.) for their helpful feedback.

\section*{Declarations}

The author declares the following competing interests.
\begin{itemize}
\item Financial competing interests: The author is self-employed as a consultant in the field that is the subject of this Review.
\item Non-financial competing interests: The author is an unpaid member of committees of the American Astronomical Society and International Astronomical Union that advocate or lobby for interests that are the subject of this Review.
\end{itemize}

\bibliography{NATASTRON-22015003A}


\end{document}